\documentclass[aps,prfluids,twocolumn,longbibliography,superscriptaddress,groupedaddress]{revtex4-2} 
\usepackage{graphicx}
\usepackage{dcolumn}
\usepackage{bm}
\usepackage[hidelinks]{hyperref}
\usepackage[version=3]{mhchem} 
\usepackage[T1]{fontenc}       
\usepackage{siunitx}
\usepackage{amssymb}
\usepackage{amsmath}

\begin{document}

\author{Pierre Lidon}
\affiliation{Robert Frederick Smith School of Chemical and Biomolecular Engineering, Cornell University, 120 Olin Hall, Ithaca, NY, USA}
\affiliation{CNRS, Solvay, LOF, UMR 5258, Univ. Bordeaux, 178 avenue du Dr. Schweitzer, F-33600 Pessac, France}
\email{pierre.lidon@u-bordeaux.fr}

\author{Etienne Perrot}
\affiliation{Robert Frederick Smith School of Chemical and Biomolecular Engineering, Cornell University, 120 Olin Hall, Ithaca, NY, USA}

\author{Abraham D. Stroock}
\affiliation{Robert Frederick Smith School of Chemical and Biomolecular Engineering, Cornell University, 120 Olin Hall, Ithaca, NY, USA}
\affiliation{Kavli Institute at Cornell for Nanoscale Science, Physical Sciences Building, Ithaca, NY, USA}
\email{ads10@cornell.edu}

\title{Non-isothermal effects on water potential measurement in a simple geometry}

\begin{abstract}
In this paper, we investigate quantitatively the coupling between gradients of temperature and of chemical or water potential under steady state conditions in the vapor phase. This coupling is important for the measurement and modeling of the dynamics of water in unsaturated environments like soils and plants. We focus on a simple non-equilibrium scenario in which a gradient of temperature exists across an air-filled gap that separates two aqueous phases with no net transfer of water. This scenario is relevant for measurements of the water potential in environmental and industrial contexts. We use a new tool, a microtensiometer, to perform these measurements. We observed variations of water potential with difference of temperature across the air gap of $-\SI{7.9(3)}{\mega\pascal\per\kelvin}$, in agreement with previous measurements. Our result is close to a first order theoretical prediction, highlighting that most of the effect comes from the variation of saturation pressure with temperature. We then show that thermodiffusion (Soret effect) coupled to natural convection could occur in our experiment and discuss how these effects could explain the small discrepancy observed between measurements and first order theoretical prediction.
\end{abstract}

\maketitle

\section{Introduction}

\subsection{Water transport in non-isothermal unsaturated media}

\begin{figure*}
\centerline{\includegraphics[width=12cm]{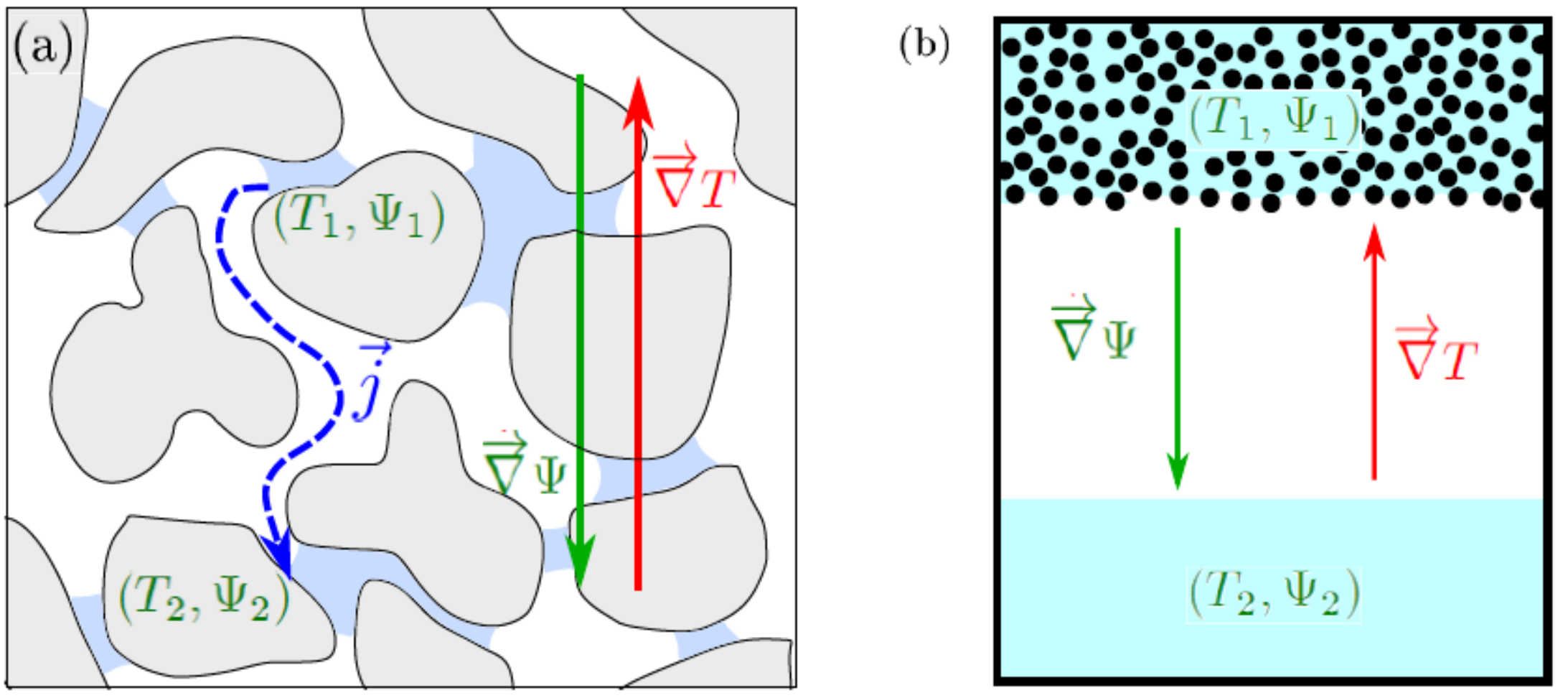}}
\caption{Context of interest. (a) Unsaturated porous medium. Within the porous matrix, pockets of capillary condensed liquid water coexist with pockets of vapor in air. In the presence of a temperature gradient, a gradient of water potential is established which drives a net water flux, $\vec{j}$, through the medium. The gradient of water potential, $\nabla \Psi$, is proportional to the gradient of temperature, $\nabla T$ (see Eq.~\eqref{eq:dPsi_dT_simple_theory}). (b) Simplified situation studied here. Aqueous phases are kept separated by a vapor gap, in a close system ensuring that there is zero net flux of mass ($\vec{j}=\vec{0}$). An imposed stationary temperature gradient, $\nabla T$, results in a steady water potential gradient, $\nabla \Psi$.  We focus on the special case in (b) in this study.\label{fig:schematics}}
\label{fig:context}
\end{figure*}

In countless natural and technological contexts, multiphase transport of a mass and heat energy occurs in the presence of gradients of both chemical potential and temperature. Many examples of this process involve water, as in unsaturated soils under arid atmospheric conditions and solar radiative forcing~\cite{scanlon_1994a,scanlon_1994b}, the drying of porous materials~\cite{scherer_1990,shokri_2015}, the cooking of foods~\cite{datta}, and the operation of air-breathing electrodes in fuel cells~\cite{weber_2014,banerjee_2015}.  As presented schematically in Fig.~\ref{fig:schematics}(a), a porous solid hosts interspersed liquid and vapor phases in contexts such as these.

In this study, we use the ``water potential'' $\Psi \, [\SI{}{\pascal}]$ which is a commonly used quantity in soil and plant science to describe water status, as water fluxes is directed from areas of high water potential towards area of low water potential. It is defined by the following equation~\cite{slatyer_1960}:
\begin{equation}
\Psi(T,P) = c_\ell [\mu(T,P) - \mu_\text{sat}(T)]
\label{eq:def_psi}
\end{equation}
\noindent where $\mu(T,P) \, [\SI{}{\joule\per\mole}]$ is the chemical potential of water in the system, $\mu_\text{sat}(T) = \mu(T,P_\text{sat}(T)) \, [\SI{}{\joule\per\mole}]$ is the chemical potential at saturation ($P_\text{sat}(T)$ being the vapor saturation pressure) and $c_\ell \, [\SI{}{\mole\per\meter\cubed}]$ is the mole density of the liquid phase. Water potential thus measures the deviation from saturation in a substance containing water: from now on, we will express it in $\SI{}{\mega\pascal}$ which provides a typical order of magnitude of water potential in our experiments as well as in field situations. In this study, we focus our attention on water, but the phenomena explored are relevant to other substances undergoing multiphase transport within a porous host matrix, as in packed bed reactors~\cite{alper,iliuta}.

The combination of various transport phenomena, local phase equilibria, and interfacial phenomena within complex geometries make the study of such processes challenging. A strong tradition of theoretical and experimental studies of multiphase transport in porous media exists in the field of soil physics. Since the foundational theoretical work on this topic by Philip and de Vries proposed~\cite{philip_1957a,philip_1957b,devries_1958,devries_1987,milly_1982}, numerous studies have attempted to characterize the various contributions to the flux of mass and heat -e.g., vapor diffusion and convection, capillary convection, surface transport, conduction- but important uncertainties remain~\cite{novak_2016,parlange_1998,vanderborght_2017}.

\subsection{Coupling between temperature and water potential gradients}

Perhaps surprisingly, we still lack clarity on one of the most basic seeming contributions to the flux of mass and energy: the transfer, across a volume of vapor , between two volumes of liquid at different temperatures and chemical potentials. Such a situation occurs between pockets of the condensed phase in realistic settings depicted in Fig.~\ref{fig:schematics}(a). Throughout this study, we focus on a simple, steady-state realization of this scenario shown in Fig.~\ref{fig:schematics}(b) in which a constant difference in temperature, $\Delta T \, [\SI{}{\kelvin}]$ and a constant heat flux, $\vec{j}_q \, [\SI{}{\watt\per\meter\squared\per\second}]$ are maintained between two condensed phases of water across air- and vapor-filled gas, with no net transfer of mass ($\vec{j}=\vec{0}$ at the boundaries of the system). We examine in detail how the steady state difference in water potential $\Delta \Psi$ depends on this difference in temperature $\Delta T$ : in what follows, we will refer to this effect as the $\Psi-T$ coupling. This dependence is important in the context of measurements of water potential in complex media such as soils, plant tissues, and food stuffs~\cite{dixon_1984,pagay_2014,black_2019}.

In this situation, the water potential $\Psi_\ell$ of the condensed phases can vary due to differences in pressure, osmolarity, and interfacial interactions with the host matrix:
\begin{equation}
\Psi_\ell = (P-P_\text{sat}(T)) - \Pi + \Psi_\text{matrix}
\label{eq:water_pot_liquid}
\end{equation}
where $P$ and $P_\text{sat}(T) \, [\SI{}{\mega\pascal}]$ are the total gas pressure and saturation partial pressure of the pure liquid, $\Pi \, [\SI{}{\mega\pascal}]$ is the osmotic pressure due to the presence of solutes, and $\Psi_\text{matrix} \, [\SI{}{\mega\pascal}]$ is the contribution of specific interactions with the matrix (e.g., van der Waals interactions of an adsorbed film). In the gas phase, the water potential of the vapor, $\Psi_\text{g}$ has the form:
\begin{equation}
\Psi_\text{g}(T,P) = c_\ell R T \ln{\left(\frac{P_\mathrm{H_2O}}{P_\text{sat}(T)} \right)}
\label{eq:water_pot_ideal_gas}
\end{equation}
\noindent where $P_\mathrm{H_2O}$ is the partial pressure of water in the gas phase, and $P_\mathrm{H_2O}/P_\text{sat}$ is the relative humidity of the vapor. By introducing the mole fraction $x$ of water in the gas phase and assuming the gas mixture behaves ideally, the partial pressure can be written as $P_\mathrm{H_2O}=xP$. At the interfaces between vapor and the condensed phases, we assume local equilibrium, which imposes that $\Psi_\ell = \Psi_\text{g}$.  Therefore, the non-equilibrium processes that drive the $\Psi-T$ coupling should occur within the gas phase if we neglect all other paths (e.g., surface diffusion along solid boundaries) connecting the two reservoirs. 

\subsection{A first-order model of the $\Psi-T$ coupling}
\label{subsec:first_order_model}
 	
Even in this highly simplified context, a variety of physical phenomena can come into play in defining the dependence of water potentials on the disequilibrium in temperature. These phenomena include: the temperature-dependences embedded in the local phase equilibria through Eqs.~\eqref{eq:water_pot_liquid} and~\eqref{eq:water_pot_ideal_gas}; the temperature-dependences of the density and local concentrations in the mixed gas phase through which heat and mass can be exchanged between the two reservoirs; and molecular diffusion, thermodiffusion, and convection driven by variations in composition, temperature, and density within the gas. In this section, we develop a first order model accounting only for this latter effect; we discuss the other possible contributions in the Discussion, after presenting our experimental observations.

Taking the derivative of Eq.~\eqref{eq:water_pot_ideal_gas} with respect to temperature, we obtain
\begin{equation}
\left( \frac{\mathrm{d} \Psi}{\mathrm{d} T} \right)^0 = c_\ell R \left( \frac{\Psi}{c_\ell R T} - \frac{T}{P_\text{sat}} \frac{\mathrm{d} P_\text{sat}}{\mathrm{d} T}\right).
\label{eq:dPsi_dT_simple_theory}
\end{equation}
\noindent We use the superscript $0$ to refer to this simple model, where total gas pressure $P$ and partial vapor pressure of water $P_\mathrm{H_2O}$ (or mole fraction $x=P_\mathrm{H_2O}/P$) are assumed to be homogeneous in the gas-filled gap (see Fig.~\ref{fig:context}(b)). In our experiments and many common contexts, the phases are near saturation such that $|\Psi| \ll c_\ell RT \sim \SI{120}{\mega\pascal}$ so the first term in this equation is negligible. Thus, the variation of water potential with temperature, is essentially due to the dependency of the saturation pressure. For the liquid-gas equilibrium, this variation is classically described by the Clausius-Clapeyron equation:
\begin{equation}
\frac{\mathrm{d} P_\text{sat}}{\mathrm{d} T} \simeq \frac{P_\text{sat} \ell_\text{vap}}{RT^2}
\end{equation}
\noindent where $\ell_\text{vap} \, [\SI{}{\joule\per\mole}]$ is the latent heat of evaporation of water. By using values from the IAPWS equation of state~\cite{wagner_2002} and for a temperature $T=\SI{21}{\celsius}$ corresponding to our experiment, we obtain that
\begin{equation}
\left( \frac{\mathrm{d} \Psi}{\mathrm{d} T} \right)^0 = - \SI{8.33}{\mega\pascal\per\kelvin}.
\label{eq:dPsi_dT_value_theory}
\end{equation}
\noindent This coupling of water potential and temperature creates the potential for a substantial driving force for water transport throughout the soil-plant-atmosphere continuum due to ambient gradients of temperature or those generated by solar radiative forcing~\cite{jarvis_1986,wang_2012,rockwell_2014}. It could also be exploited for heat management technologies~\cite{narayana_2012}. This coupling between water potential and temperature is actually general for any phase equilibrium and can be described by similar arguments~\cite{koopmans_1966,thesis_Robin}.

\subsection{Experimental investigations with psychrometers}
\label{subsec:psychrometers}

This $\Psi-T$ coupling is the basis of the measurement of water potential by thermocouple-based psychrometers~\cite{spanner_1951,richards_1958}. In this original approach, a single thermocouple is held in an air-filled cavity adjacent to the sample of interest. A first measurement of the air temperature provides the wet-bulb temperature. Then, an applied voltage would cool the junction by the Peltier effect until condensation occurred. The condensation temperature provides the dew point temperature. After calibration of the thermocouple junction, the measured dew point can be associated with the water potential of the system, as justified by Eq.~\eqref{eq:dPsi_dT_simple_theory}. A detailed sensitivity analysis of thermocouple-based psychrometers has been proposed~\cite{rawlins_1966,peck_1968}. This original design with a single thermocouple assumes that the air and the sample are at thermal equilibrium and it is highly sensitive to Psi-T coupling in the presence of thermal gradients. 

Later, Dixon and other introduced a psychrometer design with two thermocouples, one in contact with the sample to measure its temperature and one operated as described in the previous paragraph~\cite{dixon_1984,wullschleger_1988}. In this approach, one of the thermocouples is used to measure the temperature of a wetted sample of known water potential while the second thermocouple is separated from the first (and the matrix) by an air-filled gap. This second thermocouple is used to measure the air temperature (the ``dry-bulb temperature''), and then cooled until condensation occurs to define the dew point in the air (the ``wet-bulb temperature''). This operation creates, transiently, the scenario depicted in Fig.~\ref{fig:schematics}(b). 

They investigated this $\Psi-T$ coupling in an applied perspective of correcting systematic uncertainties in water potential measurements due to a temperature difference between the sample under study and the wet-bulb. Psychrometers indeed measure water potential of the vapor surrounding the wet bulb: if it is not in thermal equilibrium with the sample of interest, the measured water potential will be biased due to the $\Psi-T$ coupling, and the prediction of Eq.~\eqref{eq:dPsi_dT_value_theory} shows that the associated error is significant. For example, a temperature difference of $\SI{0.1}{\kelvin}$ leads to a systematic error of $\SI{0.8}{\mega\pascal}$ on the measured water potential, which is significant for applications in soils and leaves for which the usual order of magnitude of water potentials is a few $\SI{}{\mega\pascal}$. They thus purposely imposed a controlled temperature gradient in the measuring chamber : in a first study they estimated a variation of $-\SI{7.77}{\mega\pascal\per\kelvin}$~\cite{dixon_1984} and later, in a more systematic study with an improved setup~\cite{wullschleger_1988}, obtained values ranging between $-\SI{7.53}{\mega\pascal\per\kelvin}$ and $-\SI{7.84}{\mega\pascal\per\kelvin}$. However, they did not give any detail on measurement uncertainties (in particular, they did not discuss the sensitivity analysis proposed by Peck~\cite{peck_1968}) and they only investigated small temperature variations of about $\SI{0.1}{\kelvin}$; their measurements appear to significantly differ from the theoretical prediction of Eq.~\eqref{eq:dPsi_dT_value_theory} and without uncertainty analysis it is not possible to discuss the relevance of this discrepancy.

\subsection{Model situation of this study}

As it appears from the previous discussion, despite its importance both for water transport in unsaturated porous media and for measurement of water potential, there is no direct experimental confirmation of the theoretical prediction of Eq.~\ref{eq:dPsi_dT_simple_theory}. In particular, existing studies do not perform quantitative comparison between measurements and theory, as would require a careful uncertainty analysis, and they mostly rely on measurements with thermocouple-based psychrometers which also rely on exploiting the $\Psi-T$ coupling.

In this study, we present an experimental measurement of the $\Psi-T$ coupling using a distinct sensing tool, a microtensiometer, with the goals of extending the measurement to a larger range of temperature and providing a thorough assessment of uncertainties. The microtensiometer (see Fig.~\ref{fig:tensiometer}) exploits distinct physics relative to the psychrometer to transduce water potential into an electronic signal and allowed us to perform measurements at true steady states and independently of any assumption on the $\Psi-T$ coupling. We worked in an experimental system (see Fig.~\ref{fig:setup}) designed to recreate the simple scenario depicted in Fig.~\ref{fig:schematics}(b) as closely as possible by providing precise control of temperature gradients, isolating the contributions of processes associated with the gas phase and vapor-liquid equilibrium, and independently varying the isothermal difference in water potential.  With this approach, we provide the most complete experimental characterization of the $\Psi-T$ coupling of which we are aware. We find a constant of proportionality $-\SI{7.9(3)}{\mega\pascal\per\kelvin}$ that is compatible, within uncertainty, with that reported previously, but in limited agreement with the theoretical prediction in Eq.~\eqref{eq:dPsi_dT_value_theory}. This discrepancy motivates a discussion of other contributions to the $\Psi-T$ coupling and the proposal of future studies.

\section{Materials and methods}
\label{sec:mat_met}

\subsection{Microtensiometer}
\label{subsec:microTM}

Despite its importance, water potential is a challenging quantity to measure reliably and in situ~\cite{loescher_2007}. The aforementioned thermocouple-based psychrometers have been the most relevant tool available for ex-situ water potential measurement as they provide appropriate sensitivity to water potential over a useful range~\cite{scanlon_1994b}. Nonetheless, the use of psychrometers has been limited by challenges in their operation (e.g., due to microvolt-scale, transient signals) and in evaluating local temperature gradients with the air gap that must separate the sensing junction from the sample~\cite{dixon_1984}.

\begin{figure*}
\centerline{\includegraphics[width=\textwidth]{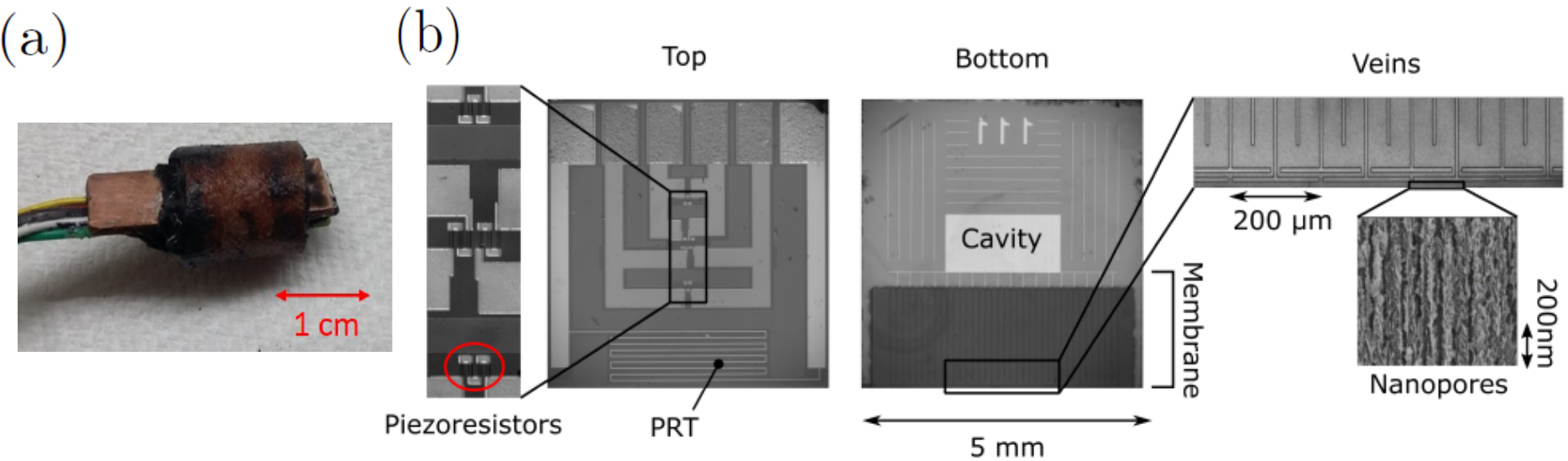}}
\caption{Pictures of the tensiometer. (a) Global view of a packaged tensiometer with a copper strip. See Fig.~\ref{fig:setup}(a,b) for a schematic diagram of the packaged device. (b) Details of the tensiometer. The top view shows the piezoresistors (an example circled in red) that constitute the strain gauge to measure deformations of the diaphragm and the platinum wire used as PRT. The bottom view shows the cavity containing bulk water and the nanoporous membrane (dark on the picture) connecting it with the outside. Microfluidic veins are etched in the membrane without linking directly the cavity and the outside in order to decrease response time of the device. Image of the nanopores has been acquired by Scanning Electron Microscopy and have average pore radius $r_\text{p}=\SI{1.7}{\nano\meter}$. Details on microfabrication are given in the text. Image credit (b): Antoine Robin.}
\label{fig:tensiometer}
\end{figure*}

A few years ago, our group introduced a microfluidic device, called the microtensiometer and presented in Fig.~\ref{fig:tensiometer}, that allows for the local (the sensor being smaller than $\sim \SI{1}{\centi\meter}$) and continuous (with a response time $\sim  \SI{1}{\minute}$) measurement of water potentials down to $\sim \SI{-20}{\mega\pascal}$~\cite{pagay_2014} relevant to conditions commonly encountered in soils and plants. Such a sensor complements the psychrometers in that it senses water potential by a distinct physical mechanism: it measures pressure (tension) in pure liquid water in equilibrium with the phase of interest. We recently proved it can provide users with simple and reliable measurements of water potential both ex situ and in situ for food products~\cite{black_2019}. It thus opens an opportunity to monitor water status within soils and plants and gives a new platform to explore water fluxes in unsaturated media.

We now recall the characteristics of the microtensiometer; further details on the fabrication has been published elsewhere \cite{black_2019}. The microtensiometer is a microfluidic tool designed to measure water potential in unsaturated phases. The size of the chip is of $\sim \SI{5}{\milli\meter}$ on a side and $\sim \SI{1}{\milli\meter}$ thick; when packaged, the size of the sensor is of order of centimeter. The internal liquid equilibrates with the external environment via the exposed edge of the porous silicon membrane; this exposed edge is $\sim \SI{5}{\micro\meter}$ thick and $\sim \SI{4}{\milli\meter}$ wide. This small size allows for local measurements.

Fig.~\ref{fig:tensiometer} presents pictures of the microtensiometers used in this study; the design is presented in more detail in~\cite{pagay_2014}. The key elements of this design are:
\begin{enumerate}
\item An internal cavity ($\SI{1}{\milli\meter} \times \SI{2.5}{\milli\meter} \times \SI{3}{\micro\meter}$) formed by dry etching in a single crystal silicon wafer $\langle 111 \rangle$ and filled with liquid water in a pressure chamber ($\sim \SI{3.5}{\mega\pascal}$ during $\sim \SI{10}{\hour}$) before use.
\item A strain gauge formed of a piezoresistive polysilicon on the opposite side of the silicon wafer from the cavity. The piezoresistors form a Wheatstone bridge.
\item A layer of porous silicon ($\SI{5}{\micro\meter}$-thick) formed by anodization on the side of the wafer that contains the cavity. This layer is filled with liquid water and provides a connected path of nanoscopic pores (average pore radius $r_\text{p} \simeq \SI{1.7(4)}{\nano\meter}$) from the cavity to one edge of the device. We have characterized the hydraulic and wetting properties of this material previously \cite{vincent_2016}.
\item Microfluidic veins, also filled with liquid water, partially spanned the distance between the cavity and the edge within porosified region. These veins increase the permeability of the path between the cavity and the edge.
\item A glass wafer anodically bonded to the side of the wafer with the cavity and veins.
\item A platinum resistance thermometer (PRT) formed by a thin platinum wire deposited on the backside of the wafer, over the porous membrane top side of the silicon. It provides a local measurement of the sensor's temperature.
\end{enumerate}

Under isothermal conditions, corresponding to $T_1=T_2$ in the scenario detailed in Fig.~\ref{fig:setup}(b), both thermal and mass transfer equilibria are reached: the temperature of the pure liquid in the cavity, $T_1$, equals that of the reference solution, $T_2$, and the water potentials of the internal liquid, $\Psi_1$, the vapor phase, $\Psi_\text{vap}$, and the reference solution, $\Psi_2$, are equal. Given that the water potential of the bulk, pure liquid water $\Psi_1 = P_\ell - P_0$ is just the difference between its pressure $P_\ell$ and this of standard state $P_0$ and the water potential of the vapor is given by Eqs.~(\ref{eq:def_psi}) and~(\ref{eq:water_pot_ideal_gas}), we have:
\begin{equation}
\Psi_1 = P_\ell - P_0 = c_\ell R T \ln{\left(\frac{P}{P_\text{sat}(T)}\right)} = \Psi_\text{vap} = \Psi_2,
\label{eq:kelvin}
\end{equation}
\noindent where $P_\ell$ is the pressure and $c_\ell$ the molar density of the liquid phase. This equilibrium is allowed by the curvature of the liquid/gas menisci at the mouth of the nanopores in the membrane; this curvature causes a reduction of the pressure of the liquid in the cavity, as shown in Fig.~\ref{fig:setup}(c), given by the Laplace law 
\begin{equation}
P_\ell = P_0 - \frac{2\gamma \cos\theta}{r_\text{p}}
\label{eq:laplace}
\end{equation}
\noindent where $\gamma \, [\SI{}{\joule\per\meter\squared}]$ is the liquid/vapor surface tension of air, $\theta$ is the contact angle of the meniscus and $r_\text{p}$ is the pore radius ($r=r_\text{p}/\cos\theta)$ being the radius of curvature of the meniscus). Equilibrium can be sustained as long as the contact angle $\theta$ of the meniscus at the pore mouth is smaller than the receding angle. Previous studies by our lab indicate that the Kelvin-Laplace effect captured by Eqs.~\eqref{eq:kelvin} and~\eqref{eq:laplace} accurately describes the behavior of water in the porous silicon used in the tensiometer~\cite{vincent_2016}. Nonetheless, the precise character of the thermodynamic interactions of water within the pores of membrane do not affect the measurements performed here: at equilibrium, the pressure (a state variable) in the pure, bulk liquid in the internal cavity depends only on the state of the bulk, external phase (vapor or pure liquid) at the external surface of the membrane.

The internal cavity was bounded by a diaphragm of plain silicon; the pressure difference between the liquid in the cavity and the external atmosphere deforms this diaphragm. This deformation was measured with a strain gauge made of a Wheatstone bridge of piezoresistors deposited on the diaphragm (see Fig.~\ref{fig:tensiometer}(b)) and supplied with a fixed applied voltage of $\SI{0.4}{\volt}$. Previous work showed that the relationship between the voltage $U$ across the Wheastone bridge and water potential $\Psi$ of the water in the cavity is linear over the whole range of water potential accessible to the tensiometer~\cite{pagay_2014,black_2019,thesis_Santiago}. After a calibration procedure, that will be described in section~\ref{subsec:calibration}, it was thus possible to obtain water potential from the voltage across the bridge. Based on Eq.~\eqref{eq:kelvin}, the calibrated response of the tensiometer provides the water potential of the phase with which the microtensiometer is in equilibrium.

We calibrated the PRT in temperature against a commercial PRT (Omega) by placing them together in a water bath and changing its temperature stepwise. In this situation, the cavity and membranes were filled with liquid water and the voltage from the Wheatstone bridge was equal to its offset $U_0$, corresponding to saturated conditions. We observe that $U_0$ depended slightly on temperature: this effect induced a small systematic uncertainty; it was negligible as compared to other sources of uncertainty discussed in a later section.

\subsection{Experimental setup}
\label{subsec:setup}

\begin{figure*}
\centerline{\includegraphics[scale=1]{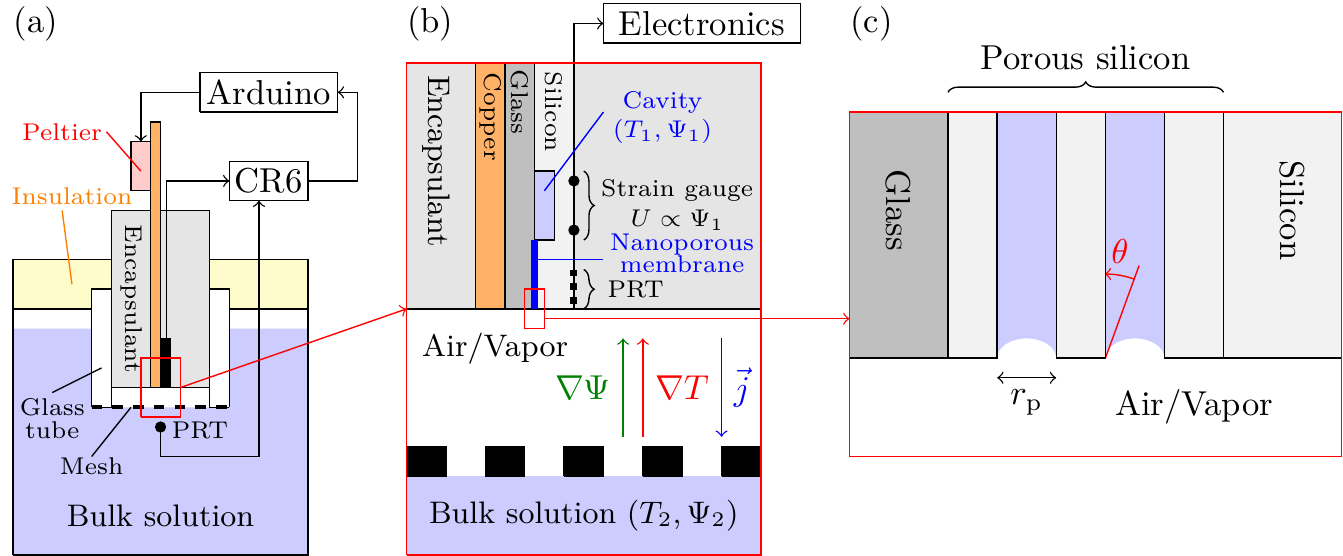}}
\caption{Schematic diagrams of the experimental setup (not on scale). (a) Cross-sectional view of the experimental setup. The tensiometer (in black) is glued to a copper strip (in orange) and packaged in urethane (in gray). The device is capped by a glass tube (in white) closed by a porous hydrophobic mesh (dashed line) and dipped in an osmotic solution of known water potential (in blue) through a thermal insulation (in light yellow). Temperature of the tensiometer is controlled by a Peltier module (in pink) attached at the emerging end of the copper strip and powered through an Arduino-based feedback loop imposing a constant temperature difference with the solution. (b) Expanded view of the vapor gap. While the temperature of the solution, $T_2$, is imposed by a thermostat, that of the tensiometer, $T_1$, is controlled with a Peltier element connected to the copper strip on which the tensiometer is glued with thermal paste. When a temperature difference is imposed, there is a difference between the water potential measured by the tensiometer, $\Psi_1$, and that of the solution, $\Psi_2$. The water potential in the cavity is obtained by measuring the deformation of the diaphragm with the strain gauge, giving a voltage $U$ proportional to $\Psi_1$. (c) Expanded view of the nanoporous membrane. When in contact with an unsaturated vapor, curvature of the menisci at the pore mouth decreases the pressure of the pore liquid to achieve mechanical equilibrium.}
\label{fig:setup}
\end{figure*}

The global view of the experimental system is depicted on Fig.~\ref{fig:setup}(a). The microtensiometer was glued to a copper strip on its glass slide, as is visible in the expanded view in Fig.~\ref{fig:setup}(b). We placed the sensor bound to the copper strip in a tube formed of resin-impregnated cardboard. The gaps around the device were filled with a curable urethane resin (UR5041, Electrolube) such that at one end the nanoporous membrane slightly protruded from the tube. In this package, the diaphragm of the tensiometer was protected from the resin so that, in the final device, it was covered by small air pocket. This air pocket avoided deformations of the diaphragm due to thermal expansion of the urethane. We capped the end of the device with a glass tube (radius $R_0 \simeq \SI{0.5}{\centi\meter}$ and height $h \simeq \SI{1}{\centi\meter}$) closed at the other extremity by a hydrophobic polymer mesh (Mo-Flow Ventilation, mesh size $\sim \SI{0.5}{\milli\meter}$) and dipped it in a solution of known water potential $\Psi_\text{ref}$. This hydrophobic mesh allowed for exchange of vapor but excluded the entry of liquid. The liquid water in the microtensiometer thus equilibrated with the solution through the vapor in the tube. This gas-filled tube volume (see Fig.~\ref{fig:setup}(b)) recreates the geometry of interest from Fig.~\ref{fig:context}(b).  

We controlled the temperature of the reference solution by submerging it in a temperature-controlled water bath, and we measured the temperature of the solution with a commercial PRT placed as closely as possible to the hydrophobic mesh. The bath was closed by a plate of insulating foam in which we drilled a tight hole to insert the probe assembly with the copper strip protruding. We attached a Peltier module (Digikey Electronics, 1681-1028-ND) to the exposed end of the copper strip, to control the temperature of the tensiometer.

A datalogger (CR6, Campbell Research Scientific) acquired data from the tensiometer and the commercial PRT in the solution. It then delivered a voltage proportional to the difference, $\Delta T^*=T_1-T_2$, between the temperature of the tensiometer $T_1$ and that of the solution $T_2$ to an Arduino board. A PID controller programmed on the Arduino switched the power supply of the Peltier module in order to impose a target temperature difference between the tensiometer and the reference solution. At steady state, this feedback loop provided a temperature difference with fluctuations of order $\SI{0.05}{\celsius}$. This fluctuation constituted a random source of uncertainty on the temperature difference that was smaller than the systematic uncertainty from the PRT calibration.

Finally, this whole setup allowed us to measure the change in water potential of the liquid water in the cavity of the tensiometer for different values of the temperature difference across the vapor gap in the glass tube and for different reference water potentials. In the following, we consider experiments during which we imposed steps of measured temperature difference, $\Delta T^*$, by keeping the solution at constant temperature and modifying the temperature of the tensiometer with Peltier module. We monitored the evolution of voltage across the Wheastone bridge continuously. In general, steps of $\sim \SI{20}{\minute}$ were long enough to allow for equilibration of the device.



\section{Results}
\label{sec:results}

\subsection{Calibration of the tensiometer}
\label{subsec:calibration}

As previously explained, equilibrium between liquid water in the cavity and water vapor surrounding the nanoporous membrane, described by Eq.~\eqref{eq:kelvin}, decreases the pressure in the cavity and deforms the diaphragm of the tensiometer. In this paragraph, we explain the calibration procedure of the sensor, allowing to estimate the water potential $\Psi$ of the vapor at the interface of the nanoporous membrane from the voltage $U$ measured across the Wheastone bridge. 

\subsubsection{Measurement in water}

\begin{figure}
\centerline{\includegraphics{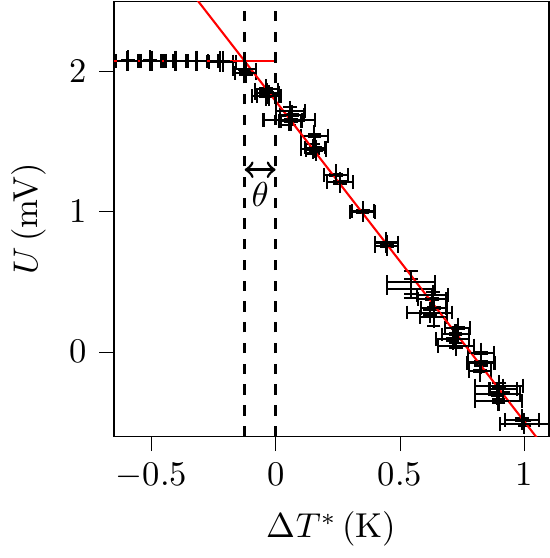}}
\caption{Evolution of the voltage $U$ across the Wheastone bridge of the tensiometer with the measured difference of temperature, $\Delta T^*$, between the tensiometer and pure liquid water. The red lines are linear regressions to determine the offset between the measured difference, $\Delta T^*$, and the true difference, $\Delta T = \Delta T^* + \theta$.}
\label{fig:exp_water}
\end{figure}

First, we used deionized water as the reference solution: by definition (Eq.~(\ref{eq:def_psi})), water potential of pure water is zero at any temperature and atmospheric pressure and $\Psi_\text{ref}=0$. The measurements are displayed on Fig.~\ref{fig:exp_water}.

As seen in Fig.~\ref{fig:exp_water}, the obtained variations of $U$ with the measured difference of temperature, $\Delta T^*$, can be divided into two parts. When temperature of the tensiometer was lower than that of the solution (corresponding to negative values of $\Delta T^*$), water vapor condensed on the nanoporous membrane (corresponding to positive values of water potential $\Psi$). As the membrane is hydrophilic, water then spreads; no deformation of the diaphragm was induced and the bridge voltage plateaued at its saturation value $U_0$. On the contrary, when the tensiometer was hotter than the solution (corresponding to positive values of $\Delta T^*$), water vapor in the air gap was unsaturated and the voltage $U$ across the Wheastone bridge evolved linearly with temperature difference.

We note, though, that the plateau voltage appeared for $\Delta T^* < 0$, while it should be reached under isothermal conditions, i.e. at $\Delta T^* =0$. This is the sign of a systematic offset $\theta = \Delta T - \Delta T^* = \SI{0.13(1)}{\celsius}$ between our temperature measurement $\Delta T^*$ and the actual temperature difference $\Delta T$ across the air gap. We determined this offset, with a good precision, by taking the intersection between the abscissa axis, $\Psi=0$, and a linear fit over the points of non-zero water potential in plots as in Fig.~\ref{fig:exp_water}.

This offset corresponds to a resistance $\sim \SI{0.9}{\ohm}$ for the PRT, which is smaller than the numerical resolution of our resistance measurements during temperature calibration. It could also stem from variation of contact resistances in the tensiometer which could have changed during manipulation between the temperature calibration and the experiment. As a consequence, we use this offset $\theta$ to correct our temperature measurements and estimate the temperature difference across the vapor gap by $\Delta T = \Delta T^* + \theta$. The uncertainty on temperature difference $\Delta T$ is mostly due to the uncertainty on the estimation of $\theta$.

\subsubsection{Measurements in osmotic solutions}

\begin{figure}
\centerline{\includegraphics{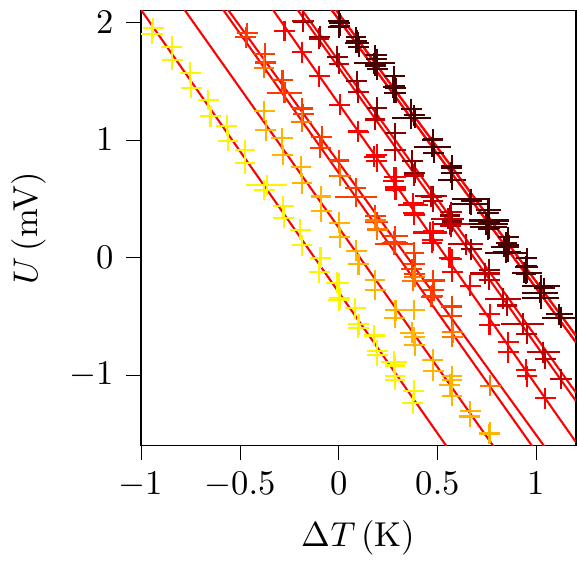}}
\caption{Variation of bridge voltage $U$ as a function of the (corrected) temperature difference $\Delta T$ across the air gap for several solutions of different water potential $\Psi_\text{ref}$ (pure water $\Psi_\text{ref}=\SI{0}{\mega\pascal}$; urea solutions $\Psi_\text{ref}=-1.16$, $-3.10$, $-3.94$, $-4.06$, $-6.13$ and $\SI{-8.88}{\mega\pascal}$; PEG solution $\Psi_\text{ref}=\SI{-1.2}{\mega\pascal}$). Lighter colors correspond to increasing absolute water potential $|\Psi_\text{ref}|$. Lines are linear fit.}
\label{fig:exp_all_data}
\end{figure}

We repeated this experiment with different solutions of solutes (sodium chloride, urea, PEG) in order to have reference solutions of varying water potentials. The reference water potential, $\Psi_\text{ref}$, of these solutions was determined by using a chilled mirror hygrometer (WP4C, Meter Group). As we worked in dilute solutions, water potential of the solution did not vary strongly with temperature and in most experiments the temperature of the water bath was kept at $\SI{21}{\celsius}$, at which the measurement of reference potential was performed. For each reference solution, we measured the voltage $U$ across the Wheastone bridge for varying temperature differences $\Delta T$; results are displayed on Fig.~\ref{fig:exp_all_data}.

Results were qualitatively similar to what we observed with water (Fig.~\ref{fig:exp_water}), except that the voltage plateaus (not displayed on Fig.~\ref{fig:exp_water}) occur for a tensiometer colder than the solution. This shift is due to the fact that water potential of the solution in isothermal conditions $\Psi_\text{ref}$ was non-zero and measurements by the tensiometer are possible as long as $\Psi \leq 0$. 

\begin{figure}
\centerline{\includegraphics{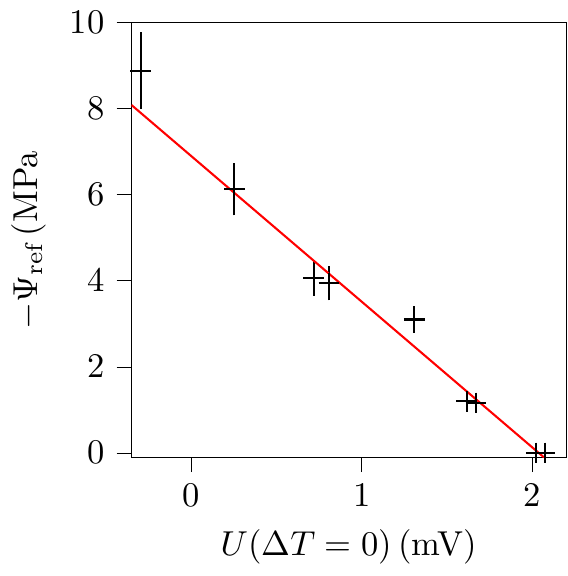}}
\caption{Absolute water potential of the solution $-\Psi_\text{ref}$, determined by a chilled mirror hygrometer, as a function of the bridge voltage $U(\Delta T=0)$ measured by the tensiometer in isothermal conditions. Red line corresponds to a linear fit according to Eq.~\eqref{eq:calibration_wp}.}
\label{fig:osmotic_calibration}
\end{figure}

As displayed on Fig.~\ref{fig:exp_all_data}, we performed linear fits of the the $U(\Delta T)$ curves for the different solutions, and extracted the isothermal voltage $U(\Delta T = 0)$ for every reference. The result is displayed on Fig.~\ref{fig:osmotic_calibration}.

As expected, there is a linear relationship between the isothermal voltage $U(\Delta T = 0)$ and water potential of the solution $\Psi_\text{ref}$ and our data can be fitted by the following equation:
\begin{equation}
-\Psi_\text{ref} = \alpha U(\Delta T=0) + \beta
\label{eq:calibration_wp}
\end{equation}
\noindent with $\alpha = \SI{-3.4(2)}{\mega\pascal\per\milli\volt}$ and $\beta = \SI{8.0(4)}{\mega\pascal}$. By using this equation, we were able to convert the measured voltage $U$ across the Wheastone bridge under any condition into the water potential $\Psi$ of the vapor at the nanoporous membrane.

\subsubsection{Measurement of the $\Psi-T$ coupling}

\begin{figure}
\centerline{\includegraphics{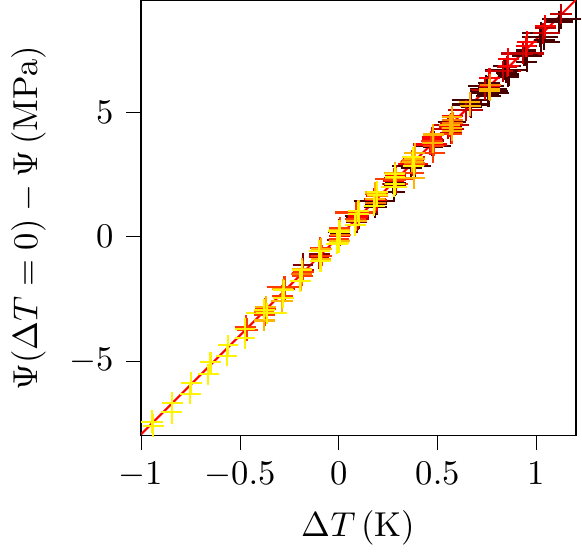}}
\caption{Variation of water potential measured by the tensiometer relative to the value in isothermal conditions, $\Psi_\text{ref} - \Psi$, for several solutions of different $\Psi_\text{ref}$ (pure water $\Psi_\text{ref}=\SI{0}{\mega\pascal}$; urea solutions $\Psi_\text{ref}=-1.16$, $-3.10$, $-3.94$, $-4.06$, $-6.13$ and $-\SI{8.88}{\mega\pascal}$; PEG solution $\Psi_\text{ref}=-\SI{1.2}{\mega\pascal}$). Lighter colors correspond to increasing absolute water potential $|\Psi_\text{ref}|$. The red line is a linear fit.}
\label{fig:exp_urea_mastercurve}
\end{figure}

Finally, we aggregated all data on Fig.~\ref{fig:exp_urea_mastercurve} by plotting the variation of water potential $\Psi(\Delta T=0)-\Psi = \alpha(U(\Delta T=0) - U)$ of the vapor at the top of the air gap with respect to its value under isothermal conditions as a function of the temperature difference $\Delta T$ across the vapor gap.

We observe that, independently of the reference value of water potential and of the nature of the reference solution, water potential decreased linearly with the temperature difference with a constant slope
\begin{equation}
\left. \frac{\mathrm{d} \Psi}{\mathrm{d} T} \right|_\text{exp} = -\SI{7.9(3)}{\mega\pascal\per\kelvin}.
\label{eq:result}
\end{equation}

We performed all experiments shown in Fig.~\ref{fig:exp_urea_mastercurve} at the same bath temperature, $T_\text{bath}=\SI{21}{\celsius}$. We however reproduced the experiment for different bath temperatures between $T_\text{bath}=\SI{15}{\celsius}$ to $T_\text{bath}=\SI{30}{\celsius}$ (data not shown): we observed a trend of a decreasing (absolute) slope with increasing temperature, but the differences in slope were within our uncertainty. This trend is consistent with our simple prediction (Eq.~\eqref{eq:dPsi_dT_simple_theory}) that predicts a range of slope from $-\SI{8.56}{\mega\pascal\per\kelvin}$ to $-\SI{8.00}{\mega\pascal\per\kelvin}$ between $\SI{15}{\celsius}$ and $\SI{30}{\celsius}$.

\section{Discussion}

\subsection{Confrontation of our results with literature and model}

As already discussed, the question of the coupling between gradients of temperature and water potential is of crucial importance in psychrometry and has been mostly studied in this context by Dixon et al., by using their own design of psychrometer. In a first paper \cite{dixon_1984}, they observe a deviation between water potential of plant tissues measured with a mechanical technique (the Scholander Pressure Chamber~\cite{scholander_1965}) and measurements with their psychrometer: these deviations are related to the temperature difference between the sample and the psychrometer and they report a correction factor of $-\SI{7.77}{\mega\pascal\per\kelvin}$ at $T=\SI{25}{\celsius}$. In a later article \cite{wullschleger_1988}, they investigate this effect more systematically by using salt solutions of calibrated water potential as a reference and controlling more carefully the temperature gradient. In this case, they report a variation between $-\SI{7.53}{\mega\pascal\per\kelvin}$ and $-\SI{7.84}{\mega\pascal\per\kelvin}$ at $T=\SI{25}{\celsius}$. 

These values are compatible with our measurements (Eq.~\eqref{eq:result}) if we account for our measurement uncertainties. However, comparison is hard to make as they do not discuss uncertainties. Moreover, they are restricted to small temperature differences below $\SI{0.1}{\kelvin}$ and small tensions below $\SI{5}{\mega\pascal}$. Our study thus significantly extends the range of measurement of the $\Psi-T$ coupling (with both positive and negative temperature differences over a $\SI{2}{\degree}$ range in various solutions) and includes experimental uncertainties.

The theoretical value obtained with the simple model ($-\SI{8.33}{\mega\pascal\per\kelvin}$, Eq.~\eqref{eq:dPsi_dT_simple_theory}) sits slightly outside of the experimental uncertainty of our measurements ($-\SI{7.9(3)}{\mega\pascal\per\kelvin}$, Eq.~\eqref{eq:result}). This discrepancy is small, in particular as uncertainties are hard to estimate with precision in such a measurement. However, it seems significant, in particular as previous measurements also report values which lie below the theoretical prediction. It is thus interesting to look for possible biases both in the experiment and in the model.

\subsection{Possible biases in the experiment}

In designing the setup and performing the experiments, we took great care to avoid possible systematic uncertainties in our measurements. However, it is never possible to guarantee their absence. A first possibility would be the presence of impurities at the liquid/vapor interfaces\cite{koopmans_1966}, but this effect is likely to be negligible as we took care to work with deionized water and did not observe any systematic evolution of our measurements when repeating the experiment.

Precise temperature control and measurement are an experimental challenge and, due to the large variations of water potential caused by the $\Psi-T$ coupling, small errors on measured temperature difference lead to significant errors on the $\Psi-T$ coupling. Such errors can arise between the temperatures at the ends of the vapor gap, and these at the location of the thermometers. We already described the correction of the observed offset $\theta = \Delta T - \Delta T^*$ between the measured and actual temperature differences.

However, there is another possible source of error in temperature measurement: due to finite thermal conductivity of water and porous silicon, we cannot rule out the existence of temperature gradients between the on-chip PRT and the cavity in the tensiometer, and between the reference PRT and the liquid-vapor interface in the solution. These discrepancies would cause a linear relationship $\Delta T^* = \gamma \Delta T$ with $\gamma \neq 1$ and we overall expect an affine relation ship $\Delta T^* = \gamma \Delta T + \theta$. As the PRT of the microtensiometer is placed at its tip and silicon is a good heat conductor, we do not expect significant error on the measurement of the temperature on the tensiometer side. If such an error could occur, as the PRT is further from the heat source than the cavity, it would lead to an underestimation of the temperature difference and thus to an overestimation of the value of $(\mathrm{d} \Psi /\mathrm{d} T)$. Correction of this effect would tend to give a correction toward lower values of this coefficient and thus cannot explain the observed discrepancy with the theoretical prediction.

The measurement of the temperature at the mesh side is more difficult as we cannot place precisely the thermometer exactly at the end of the vapor gap, in contact with the mesh. To control this possible uncertainty, we repeated the experiment without the mesh such that we could place the tip of the reference PRT inside the vapor gap. In this situation, we obtained exactly the same value for the slope $\mathrm{d}\Psi/\mathrm{d}T$, i.e., $-\SI{7.9}{\mega\pascal\per\kelvin}$ (data not shown). This makes us confident that the correction factor $\gamma$ is very close to $1$ and our estimation of temperature differences (corrected by the offset) were reliable.

\subsection{Refinement of the model}
\label{subsec:refinements}

Derivation of Eq.~\eqref{eq:dPsi_dT_simple_theory} relies on the assumption that both pressure and mole fraction of water are constant throughout the vapor gap. A more general relation can be obtained by differentiating Eq.~\eqref{eq:water_pot_ideal_gas}:
\begin{widetext}
\begin{equation}
\mathrm{d} \Psi = c_\ell R \left[ \frac{\Psi}{c_\ell R T} - \frac{T}{P_\text{sat}(T)} \frac{\mathrm{d} P_\text{sat}}{\mathrm{d} T} \right] \mathrm{d} T + c_\ell R T \left( \frac{\mathrm{d} P}{P} + \frac{\mathrm{d} x}{x} \right) \equiv \left(\frac{\mathrm{d} \Psi}{\mathrm{d} T}\right)^0 \mathrm{d} T + c_\ell R T \left( \frac{\mathrm{d} P}{P} + \frac{\mathrm{d} x}{x} \right).
\label{eq:psychrometric_effect_full}
\end{equation}
\end{widetext}
\noindent The first term of this equation corresponds to the first-order theory presented in section~\ref{subsec:first_order_model} (Eq.~\eqref{eq:dPsi_dT_simple_theory}). In this theory, we assumed pressure and composition remain homogeneous across the vapor gap. The second term in Eq.~\eqref{eq:psychrometric_effect_full} quantifies the corrections to the first order-theory of $\Psi-T$ coupling due to possible variations in pressure and mole fraction.

From the right hand side of Eq.~\eqref{eq:psychrometric_effect_full}, we can get an idea of the gradients of pressure and mole fraction that would need to be present to match the discrepancy between our measurement and theory. For example, for a temperature difference of $\Delta T = \SI{1}{\kelvin}$, we have:
\begin{equation}
\frac{\Delta P}{P} + \frac{\Delta x}{x} = \frac{\Delta \Psi - \Delta \Psi^0}{c_\ell RT} \sim \SI{0.3}{\percent}
\end{equation}
\noindent where $\Delta \Psi^0 = (\mathrm{d}\Psi / \mathrm{d}T)^0 \Delta T$ is the prediction from the simple theory. Consequently, moderate variations of pressure or mole fraction could explain the small discrepancy between our measurements and the simple theory; in the following, we describe possible causes of such variations and the order of magnitude of their impact on $\Psi-T$ coupling. Further details can be found in Appendix~\ref{SI:sec_details}.

\paragraph{Pressure gradients}

Pressure gradients are unlikely to explain the observed discrepancy. Hydrostatic pressure gradient on the small air column of height $h=\SI{1}{\centi\meter}$ we studied is completely negligible and, as detailed in Appendix~\ref{SI:subsec_pressure}, pressure gradients generated by possible convection flow in the air gap would require a non-realistic flow speed around $v \sim \SI{e4}{\meter\per\second}$. However, it is important to note that in porous media, due to the small hydraulic permeability, significant pressure gradients could occur from flows of moderate speed.

\paragraph{Cross-diffusion effects}

Mole fractions gradients can be directly created by temperature gradient through a cross-diffusion effect called thermodiffusion or Soret effect. This corresponds to a differential migration of species of a mixture in a temperature gradient. It was initially observed in salt solutions \cite{soret_1880} and later gaseous and liquid mixtures \cite{rahman_2014}. In gas mixtures, it can be predicted from kinetic theory \cite{chapman}, while, in liquids, its explanation remains elusive \cite{kohler_2016}. The Soret effect would result in a mass flux of water vapor $\vec{j}_T \; [\SI{}{\kilo\gram\per\meter\squared\per\second}]$ which is directly proportional to the temperature gradient. Such a flux superimposes to the regular diffusion flux given by Fick law, and equilibrium between them leads to a steady gradient of mole fraction. As detailed in Appendix~\ref{SI:subsec_Soret}, the resulting variation of mole fraction across the vapor gap is given by the equation:
\begin{equation}
\frac{\Delta x}{x} \simeq - \alpha_T \frac{\Delta T}{T}
\end{equation}
\noindent where $\alpha_T$ is a dimensionless coefficient, depending on the interactions between gas molecules.

We have not found values of the Soret coefficient $\alpha_T$ of water vapor in air at ambient temperature in the literature. It can be computed from kinetic theory but it requires a precise knowledge of the molecular details of the studied mixture~\cite{chapman}; by considering that the gaseous species are spherical with no internal structure and undergo elastic collisions, we estimated that $\alpha_T \simeq -0.20$ for a mixture of water and air at $T=\SI{21}{\celsius}$. These assumptions are rough but it has been showed that this treatment gives a correct sign and order of magnitude of the coefficient $\alpha_T$ for mixtures of water and molecular hydrogen under large temperature gradients~\cite{whalley_1951b}. 

Qualitatively, we observe that $\alpha_T <0$, which corresponds to an accumulation of water at hot places and a reduction of the measured absolute water potential with respect to the simple prediction; this is in qualitative agreement with our experiment. Quantitatively, we find that Soret effect induces a correction to the $\Psi-T$ coupling leading to $(\mathrm{d}\Psi / \mathrm{d} T) = \SI{-8.24}{\mega\pascal\per\kelvin}$: the Soret effect could have an impact of about $\SI{1}{\percent}$ on the $\Psi-T$ coupling. This contribution would not be negligible and could bring the prediction within the uncertainty of our measurements.

Another cross-diffusion effect arises in presence of a pressure gradient, called barodiffusion. In this case, the transport coefficient can be directly obtained from molar masses and the composition of the difference species, with no need for a detailed microscopic theory~\cite{landau_mecaflu}. However, as argued in the previous paragraph, pressure gradients should be very small in our experiment and barodiffusion should have a negligible impact.

\paragraph{Natural convection}

Finally, in our estimate of the impact of Soret effect, we assumed the gas was at rest. Natural convection by itself cannot generate a steady concentration gradient in the air gap. However, it can couple with Soret effect; this could lead to an enhanced separation of the species~\cite{kozlova_2016}, here corresponding to further accumulation of water around the tensiometer. Such a effect would further reduce the absolute water potential and could possibly explain the discrepancy between our measurement and the simple theory of $\Psi-T$ coupling. A full (numerical) resolution of the advection-diffusion equation for water vapor accounting for the Soret effect would be required to estimate precisely the impact of convection on the $\Psi-T$ coupling.

In most of our experiments, the temperature gradient is upwards, i.e., the gas in the tube is heated from above; this situation is stable with respect to convection and no flow should appear. In the few cases in which the temperature gradient was downwards (with osmotic solutions), temperature gradients are too small to trigger thermal convection.

However, temperature gradient is not purely vertical in our experiment. As thermal conductivity of the glass tube is significantly larger than this of the gas inside, there should be a radial temperature gradient, with temperature decreasing from the center of the tube towards the walls. Such a situation in which the temperature gradient is orthogonal to gravity, leads to convection flows without any threshold, with fluid going upwards where the temperature is higher and downwards when it is lower. In our case, we thus should have a vertical upwards flow at the center of the cell. As detailed in Appendix~\ref{SI:subsec_convection}, simple estimates lead to a maximum recirculation velocity of $v_\text{max} \simeq \SI{2e-4}{\meter\per\second}$.

\section{Conclusion and perspectives}
\label{sec:conclusion}

In this paper, we studied experimentally the evolution of water potential caused by a temperature gradient across a vapor in a closed system and obtain a value of $-\SI{7.9(3)}{\mega\pascal\per\kelvin}$ at $\SI{21}{\celsius}$ in agreement with previous experimental results. This result is close to a simple theoretical prediction, assuming the gas phase is at rest and displays no gradients of pressure and mole fraction; we conclude that most of the effect comes from the variation of saturation pressure with temperature. However, this first order prediction lies slightly outside our experimental uncertainty and previous experiments also systematically obtained values below the theory. We thus investigated the possible source of errors both in the experiment and the modeling of the situation.

While always possible, it seems that systematic biases in the experiment (in particular regarding temperature control and measurement) cannot explain the discrepancy as these effects would tend to overestimate the real effect. However, we showed that modeling of our setup is more complex than could be thought at first sight due to the possibility of thermodiffusion flux (Soret effect) and its coupling with natural convection occurring due to radial component of temperature gradient. Future experiments will improve the setup in order to decrease measurement uncertainties (in particular by using more careful calibration procedures of the microtensiometer) and should be performed with a pure water vapor gap (i.e., in an evacuated cell). In this case, there should be no Soret effect and mole fraction should be one everywhere. If we confirm that the Soret effect is at the origin of the observed discrepancy, our setup could give an interesting method to measure Soret coefficients of various molecules in gases as the microtensiometer can be operated with other fluids than water~\cite{thesis_Santiago}.

As confirmed by our work, water potential varies strongly with temperature; this highlights the necessity, well known by users of psychrometers, to ensure isothermal conditions or to apply a proper correction if it is not the case. This same caution applies to the use of our microtensiometer for the measurement of water potential in unsaturated media. Nonetheless, the design and mode of operation of the microtensiometer provides an important opportunity relative to thermocouple psychrometers: whereas the operation of psychrometers requires that the sensing element the thermocouple junction be exposed to vapor that, in turn, is exposed to the medium of interest, the nanoporous membrane of the microtensiometer can be put into direct contact with the medium of interest to favor thermal equilibrium and minimize the effect characterized here. Alternatively, for situations in which direct contact could lead to contamination, a conductive element, such as the copper strip used in this study (Fig.~\ref{fig:setup}(b)), can help maintain thermal equilibrium with the sample. Work in our lab (unpublished) with the microtensiometer embedded in the xylem tissues of woody plants indicates that good thermal contact between the tissue and the device can indeed be achieved. Finally, this strong coupling between water potential and temperature could be exploited to engineer new porous membranes for heat management \cite{narayana_2012}. 

\acknowledgments

We thank Dr. Antoine B{\'e}rut, Prof. Bernard Castaing, Dr. Pierre Fleury, Dr. Michel Fruchart, Dr. Robin Guichardaz and Dr. David Lopes Cardozo for early discussions on the modeling, Prof. John Albertson for discussions about the environmental context, and Dr. Jean-Bapsite Salmon for numerous and long discussions about Fick law and for suggesting the possibility of convection in our experiment. We are grateful to Michael Santiago, Antoine Robin and Siyu Zhu for the fabrication of the microtensiometer chips used in this study. We finally thank Prof. Michael Novak for his thorough review of the first version of the manuscript and helping to find a mistake.

The tensiometers were made in the Cornell Nanoscale Facility, an NNCI member supported by NSF grant ECCS-1542081. The authors acknowledge funding from AFOSR (FA9550-15-1-0052), NSF grant IIP 1500261, and Cornell College of Engineering.

\paragraph*{Declaration of Conflict of Interest}

Co-author A.D.S. has a financial interest in FloraPulse Co., a company that holds a license to commercialize the microtensiometer used in this study.

\appendix

\section{Details on thermodiffusion and natural convection}
\label{SI:sec_details}

In this appendix, we detail the calculations leading to the orders of magnitude of the different corrections to the first-order theory of $\Psi-T$ coupling discussed in section~\ref{subsec:refinements}.

\subsection{Possible pressure gradients}
\label{SI:subsec_pressure}

Hydrostatic pressure gradients are usually accounted for in models of soil hydrology. As gas in the vapor gap of our experiment is mostly composed of air (at the considered temperature, saturation mole fraction of water is $x_\text{sat}\sim \SI{2.5}{\percent}$) so we can assume its mass density is $\rho_\text{g} \sim \SI{1.2}{\kilo\gram\per\meter\cubed}$ which gives a pressure variation $\Delta P = \rho_\text{g} h g \sim \SI{0.1}{\pascal}$ across the vapor gap of height $h=\SI{1}{\centi\meter}$; this pressure variation is fully negligible when compared with ambient pressure.

If some convection flow occurs (see paragraph \ref{SI:subsec_convection}), it would also be coupled with pressure gradients. A simple estimate of such gradients can be obtained by using Poiseuille law, which links flow speed of a fluid of kinematic viscosity $\eta (\SI{}{\pascal})$ and pressure difference in a cylindrical pipe of height $h$ and radius $R_0$ through:
\begin{equation}
\Delta P = \frac{4\eta h}{R_0^2} v.
\end{equation}
\noindent In our case, the cell has a dimensions $2R_0 \simeq L \sim \SI{1}{\centi\meter}$ and air has a viscosity $\eta \simeq \SI{1.8e-5}{\pascal\second}$. In order to obtain a pressure variation of $\SI{0.3}{\percent}$ of the atmospheric pressure $P=\SI{1}{\bar}$, a speed $v \sim \SI{e4}{\meter\per\second}$ would be required: this flow speed in the vapor gap is not realistic.

Consequently, it is legitimate to neglect pressure gradients in our experiment. However, it is important to note that in porous media in which the permeability is small, significant pressure gradients could occur.

\subsection{Thermodiffusion}
\label{SI:subsec_Soret}

The Soret effect would result in a mass flux of water vapor $\vec{j}_T \; [\SI{}{\kilo\gram\per\meter\squared\per\second}]$ which is directly proportional to the temperature gradient and can be written as:
\begin{equation}
\vec{j}_T = - \rho_\text{g} D \alpha_T c(1-c) \frac{\overrightarrow{\nabla} T}{T}
\end{equation}
\noindent where $\rho_\text{g} \; [\SI{}{\kilo\gram\per\meter\cubed}]$ is the average mass density of the gas, $D \; [\SI{}{\meter\squared\per\second}]$ is the diffusion coefficient and $\alpha_T$ is a dimensionless coefficient~\cite{degroot}.

The total mass flux of water is thus obtained by adding Soret flux $\vec{j}_T$ to the regular Fick law $\vec{j}_\text{d} = - \rho_\text{g} D \overrightarrow{\nabla} c$:
\begin{equation}
\vec{j} = - \rho_\text{g} D \left(\overrightarrow{\nabla} c+  \alpha_T c(1-c) \frac{\overrightarrow{\nabla} T}{T}\right).
\end{equation}
\noindent As the amount of water is constant in the vapour gap, we should have $\vec{j} = \vec{0}$ so:
\begin{equation}
\overrightarrow{\nabla} c +  k_T \frac{\overrightarrow{\nabla} T}{T}=0.
\end{equation}
\noindent In the dilute regime, mole and mass fraction of water can be assimilated and we obtain:
\begin{equation}
\frac{\Delta x}{x} \simeq - \alpha_T \frac{\Delta T}{T}.
\end{equation}
\noindent The Soret effect thus induces a correction to the $\Psi-T$ coupling given by:
\begin{equation}
\mathrm{d} \Psi \simeq \left[ \left(\frac{\mathrm{d} \Psi}{\mathrm{d} T}\right)^0 - c_\ell R \alpha_T \right]\mathrm{d} T.
\end{equation}

The value of this coefficient can be predicted theoretically from kinetic theory for a binary mixture of spheres with given interaction potential \cite{chapman}. If we model the gas in the air gap as a binary mixture of water molecules and air (with average molar mass and molecular size between molecular nitrogen and oxygen) and assume collisions are elastic, it is possible to obtain an expression for $\alpha_T$ as a function of $T$, leading to a value of $\alpha_T \simeq -0.20$ in saturated air at temperature $T_0=\SI{21}{\celsius}$. We do not claim that this value is quantitative: our assumptions in kinetic theory of spherical molecules, elastic collisions, and treatment of air as a pure species with average properties are indeed dubious. However, it has been showed that this treatment gives a correct sign and order of magnitude of the coefficient $\alpha_T$ for mixtures of water and molecular hydrogen under large temperature gradients~\cite{whalley_1951b}. We thus believe that our model should give a correct order of magnitude of the impact of Soret effect.

\subsection{Hadley thermal convection}
\label{SI:subsec_convection}

In the few cases in which the temperature gradient was downwards in our experiment (with osmotic solutions), we can compute the Rayleigh number $\mathrm{Ra}=\alpha \Delta T g h^3/\nu \kappa$, where $\alpha \sim /T$ is the isobaric dilation coefficient of the gas, $\nu = \eta/\rho_\text{g}$ its kinematic viscosity and $\kappa$ its thermal diffusivity. For air, $\nu \simeq \SI{1.6e-5}{\meter\squared\per\second}$ and $\kappa \simeq \SI{2e-5}{\meter\squared\per\second}$ so for a temperature difference of $\Delta T=\SI{1}{\kelvin}$ across the vapor gap, Rayleigh number is $\mathrm{Ra}\sim 100$. Onset of convection corresponds generally to Rayleigh numbers $\mathrm{Ra} \gtrsim 1000$ so convection is very unlikely.

However, as discussed in the main text, the horizontal component of the temperature gradient could be comparable to the vertical one. In the case of a fluid contained between two infinite vertical plates separated by a distance $a$, the flow resulting from a temperature difference between the plates has an analytic expression known as Hadley flow \cite{lappa_convection_hadley}. The maximum vertical velocity in such a case is given by
\begin{equation}
v_\infty = \frac{\alpha g \Delta T}{9\sqrt{3} \nu} a^2.
\end{equation}
\noindent For a cell of finite size, the flow field is generally not known. However, for a rectangular cell of height $h$, width $a$ in the direction of the temperature gradient and infinite in the third direction, approximate solutions can be found for aspect ratio $h/a$ close to $1$ and moderate horizontal Rayleigh number (typically $\mathrm{Ra} < 1000$ corresponding to $\Delta T < \SI{10}{\kelvin}$ in our experiment) and give \cite{batchelor_1954} a maximum vertical velocity:
\begin{equation}
v_\text{max} = v_\infty  \frac{1}{h^4+a^4} \left(\frac{h}{2}\right)^4.
\end{equation}
\noindent Our cylindrical geometry is different from this situation and the horizontal temperature gradient is not constant across the vapor gap in our experiment, but this expression should give at least a correct order of magnitude of the convection speed. By taking $h = \SI{1}{\centi\meter}$ and $a=R_0=\SI{0.5}{\centi\meter}$, we obtain $v_\text{max} \simeq \SI{2e-4}{\meter\per\second}$.

\end{document}